\documentclass[12pt]{article}
\usepackage{latexsym,amsmath,amssymb}
\usepackage{graphics}
\newtheorem{proposition}{Proposition}

\begin{document}


\title{Born-Infeld black holes coupled to a massive scalar field}

\author{Ivan Zh. Stefanov$^{ 1, 2 }$ \thanks{E-mail: izhivkov@yahoo.com}\,\,\,, \,\,
     Stoytcho S. Yazadjiev$^{1 } $\thanks{E-mail: yazad@phys.uni-sofia.bg}\\
{\footnotesize  $^{1}$Dept. of Theoretical Physics,
                Faculty of Physics}\\ {\footnotesize St.Kliment Ohridski University of Sofia}\\
{\footnotesize  5, James Bourchier Blvd., 1164 Sofia, Bulgaria }\\\\[-3.mm]
  Daniela A.~Georgieva$^{2}$ \thanks{E-mail: dgeorgieva@tu-sofia.bg}\,\,\,, \,\, Michail D.~Todorov$^{2}$ \thanks{E-mail: mtod@tu-sofia.bg}
\\ [-1.mm]{\footnotesize
{$^{2}$Faculty of Applied Mathematics and Computer Science}}\\
[-1.mm] {\footnotesize {Technical University of Sofia}}\\
[-1.mm] {\footnotesize 8, Kliment Ohridski Blvd., 1000 Sofia, Bulgaria}}

\date{}

\maketitle

\begin{abstract}
Born-Infeld black holes in the Scalar-Tensor Theories of Gravity, in the case of massless scalar field, have been recently obtained \cite{SYT1, SYT2, SYT3}. The aim of the current paper is to study the effect from the inclusion of a potential for the scalar field in the theory, through a combination of analytical techniques and numerical methods. The black holes coupled to a massive scalar field have richer causal structure in comparison to the massless scalar field case. In the latter case, the black holes may have a second, inner horizon. The presence of potential for the scalar field allows the existence of extremal black holes for certain values of the mass of the scalar field  and the magnetic (electric) charge of the black hole. The linear stability against spherically symmetric perturbations is studied. Arguments in favor of the general stability of the solutions coming from the application of the ``turning point'' method are also presented.
\end{abstract}


\sloppy

\section{Introduction}
The study of black holes coupled to non-linear electrodynamics is natural since in the conditions of strong fields and strong sources, such as black holes, quantum corrections to the lagrangian of electrodynamics should be taken into account. Nonlinear electrodynamics was considered for the first time by Born and Infeld in 1934 in their attempt to obtain a finite energy density model for the electron \cite{BI}. The interest in nonlinear electrodynamics has been recently revived since such types of lagrangians appear in the low-energy limit of open strings and $D$-branes \cite{nle_str1}--\cite{L}. Nonlinear electrodynamics models coupled to gravity have been discussed  in different aspects
(see, for example,  \cite{Garcia}--\cite{SYT4} and references therein).

The properties of charged black holes coupled to Born-Infeld nonlinear electrodynamics in General Relativity have been examined in  \cite{GR1, Rasheed, Breton1, Breton2} and in the case of non-zero cosmological constant in \cite{CPW}.  The role of the derivative corrections to the properties of the Einstein-Born-Infeld black holes have been studied in \cite{TamakiJCAP}. A natural step in the study of charged black holes is to add a scalar field in the theory. Born-Infeld-dilaton (asymptotically non-flat) black holes have been reported in \cite{CG}--\cite{Sheykhi_TD_tpol}. The case of charged black holes with  massive dilaton coupled to abelian gauge field with linear lagrangian have been considered in \cite{GrH}--\cite{Tamaki}. Einstein-Born-Infeld black holes with massive dilaton have also been investigated \cite{YFBT}.

In the frame of Scalar-Tensor Theories (STT) solutions describing charged black holes with non-trivial massless scalar field in non-linear electrodynamics have been recently obtained \cite{SYT1, SYT2, SYT3}. The aim of the current paper is to extend \cite{SYT1, SYT3} by adding a potential for the scalar field that admits the presence of asymptotically flat black holes.

The solutions with massless scalar field presented in \cite{SYT1, SYT3} have a much simpler causal structure than the corresponding solutions in the Einstein-Born-Infeld (EBI) gravity. The latter have a single horizon, thus their structure resembles that of the Schwarzschild solution.  Adding a potential for the scalar field makes the causal structure richer and more difficult to study. In that case solutions with internal horizons and with degenerate event horizons may appear.

A prior intuition about the properties of the studied solutions could be obtained from other similar systems that have already been studied. An example of such systems are the charged black holes with massive dilaton coupled to linear electrodynamics. Not all techniques used in that system are applicable to our case, though. The major difference between the dilaton and the gravitational scalar in the STT is its coupling to matter\footnote{The coupling between the electromagnetic field and  scalar field in dilaton gravity and in STT is different. The difference between the two types of actions can be found in \cite{BSen}, for example.}, and in particular, to the electromagnetic field. In STT due to the specific coupling between the scalar and electromagnetic fields the source term in the field equation for the scalar field is proportional to the trace of the energy-momentum tensor of the electromagnetic field and not proportional to the lagrangian of the electromagnetic field as it is in the dilaton gravity.

The charged black holes with massive dilaton have been studied analytically in \cite{GrH} and numerical solutions have been obtained in \cite{HH}. In the first paper the conclusions about the causal structure are made. In that case black holes may have three, two or one horizons depending on the values of the parameters and the specific choice for the potential of the dilaton. The various types of extremal solutions were also studied. The authors found that two-fold and three-fold degenerate horizons may exist. The behavior of the fields near the central singularity was also obtained in that paper. In \cite{HH} approximate solution in the two limiting cases -- large black holes (with radius of the event horizon much bigger than the Compton length of the dilaton) and small black holes (with radius of the event horizon much smaller than the Compton length of the dilaton) have been obtained. Numerical solutions covering both the exterior and interior regions of the black holes are also presented.

In the current paper our aim is to study the black holes solutions in a class of STT with massive scalar field coupled to Born-Infeld nonlinear electrodynamics. In the studied region of the parameter space we exclude the possibility of the existence of a third horizon. We also study the possibility for existence of extremal black holes in different regions of the parameter space and obtain numerical solutions for them in the allowed region.

In the present paper we have commented also on the linear stability of the obtained black-hole solutions against radial perturbations. Here we follow the same scheme as for the black holes with massless scalar field \cite{SYT3}. Additional arguments in favor of the general stability of the black holes can be given also on the bases of the ``turning point'' method that we also discuss.

\section{Formulation of the problem}
The action of the studied STT is originally formulated in the so-called Jordan frame in which the scalar field is coupled to the scalar curvature and in which there is no direct coupling between the scalar field and the sources of gravity (in our case the source of gravity is the electromagnetic field). For mathematical convenience, however, as usual for the STT, we study the solutions in the conformally related Einstein frame (for more details we refer the reader to \cite{Faraoni1, Faraoni2, SYT1}). In the Einstein frame the action takes the following form
\begin{eqnarray}\label{EFA}
S= {1\over 16\pi G_{*}}\int d^4x \sqrt{-g} \left[{\cal R} -
2g^{\mu\nu}\partial_{\mu}\varphi \partial_{\nu}\varphi -
4V(\varphi)\right] \nonumber \\ + S_{m}[\Psi_{m}; {\cal
A}^{2}(\varphi)g_{\mu\nu}]
\end{eqnarray}
where ${\cal R}$ is the Ricci scalar curvature with respect to the
Einstein metric $g_{\mu\nu}$ and $G_{*}$ is the bare gravitational constant.

The action of the nonlinear electrodynamics is
\begin{equation}\label{EFNEDA}
S_{m} = {1\over 4\pi G_{*}}\int d^4x \sqrt{-g} {\cal A}^4(\varphi) L(X, Y)
\end{equation}
where
\begin{equation}
X = {{\cal A}^{-4}(\varphi)\over 4} F_{\mu\nu}{g}^{\mu\alpha} {g}^{\nu\beta} F_{\alpha\beta}, \label{X}  \,\,\,
Y = {{\cal A}^{-4}(\varphi)\over 4}  F_{\mu\nu}\left({ \star} F\right)^{\mu\nu}
\end{equation}
and ``$\star$''  stands for the Hodge dual with respect to the Einstein frame metric $g_{\mu\nu}$.

The action (\ref{EFA}) with (\ref{EFNEDA}) yields the following field equations

\begin{eqnarray}
&&{\cal R}_{\mu\nu} = 2\partial_{\mu}\varphi \partial_{\nu}\varphi +  2V(\varphi)g_{\mu\nu} -
 2\partial_{X} L(X, Y) \left(F_{\mu\beta}F_{\nu}^{\beta} -
{1\over 2}g_{\mu\nu}F_{\alpha\beta}F^{\alpha\beta} \right)  \nonumber \\
&&\hspace{2cm}-2{\cal A}^{4}(\varphi)\left[L(X,Y) -  Y\partial_{Y}L(X, Y) \right] g_{\mu\nu}, \nonumber  \\ \nonumber \\
&&\nabla_{\mu} \left[\partial_{X}L(X, Y) F^{\mu\nu} + \partial_{Y}L(X, Y) (\star F)^{\mu\nu} \right] = 0 \label{F},\\ \nonumber \\
&&\nabla_{\mu}\nabla^{\mu} \varphi = {d V(\varphi)\over d\varphi } -
4\alpha(\varphi){\cal A}^{4}(\varphi) \left[L(X,Y) -  X\partial_{X}L(X,Y) -  Y\partial_{Y}L(X, Y) \right], \nonumber
\end{eqnarray}
where $\alpha(\varphi) = {d \ln{\cal A}(\varphi)\over d\varphi}$.

In what follows we consider the truncated\footnote{Here we consider the pure magnetic case for which $Y=0$.  }
Born-Infeld  electrodynamics described by the Lagrangian

\begin{equation}
L_{BI}(X) = 2b \left( 1- \sqrt{1+ \frac{X}{b}} \right)\label{LBI}.
\end{equation}

The type of STT is determined by the specific choice of the function ${\cal A}(\varphi)$ (respectively $\alpha(\varphi)$) and $V(\varphi)$.
Here we will consider potentials $V(\varphi)$ which satisfy the conditions
\begin{equation}
\varphi \,{d V(\varphi)\over d\varphi }\geq0 \quad\hbox{and  } \quad {d V(\varphi=0)\over d\varphi }=0.\label{condPOT}
\end{equation}
In particular, for numerical calculations, we will take the potential of the scalar field in the form
\begin{equation}
V(\varphi)= {1\over 2 }m_{*}^2\varphi^2,\label{POT}
\end{equation}
where $m_{*}$ is the mass of the scalar field and has also the meaning of inverse Compton wavelength of the scalar field in the units we work. Constraints for the values of $m_{*}$ can be found in \cite{Helbig}.

In the present paper, we will be searching for solutions with regular scalar field $\varphi$ on the event horizon. We will also require that $0<{\cal A}(\varphi)<\infty$ for $r \geq r_{H}$, where $r_{H}$ is the radius of the horizon in order to ensure the regularity of the transition between the Einstein and the Jordan  conformal frames. Yet, we will consider only theories for which $\alpha(\varphi)$ has a fixed positive sign for all values of $\varphi$. The manner of investigation of solutions within theories with $\alpha(\varphi)<0$ is similar.
Theories in which the coupling function changes its sign are much more complicated (also from numerical point of view)
since in them some interesting non-perturbative effects like
``spontaneous scalarization" may appear, especially when $\alpha(\varphi)\sim\varphi$ \cite{SYT2}.

For our numerical solution we have considered theories for which the coupling function has the form
\begin{equation}
{\cal A}(\varphi)=e^{\alpha\varphi}, \\
\end{equation}
where $\alpha$ is a positive constant and in this theory $\alpha(\varphi)=\mathrm{const}\equiv\alpha$ .
We have studied the parametric space for fixed value of the coupling parameter $\alpha = 0.01$.
The conclusions for the qualitative behavior of the solutions, however, are valid for a much wider class of STT for which $\alpha(\varphi)>0$,
$\alpha(\varphi=0) \sim 10^{-4} \div 10^{-2}$ and $\beta(\varphi=0)>-4.5$ (such values are in agreement with the observations, for details see \cite{Will}), where $\beta(\varphi)={d \alpha(\varphi)\over d\varphi}$.
The exterior region solutions from the mentioned class of STT would not only be qualitatively the same but also quantitatively very close to the exterior region solutions in the Brans-Dicke theory since, as the numerical results show, $\varphi$ is very small there and higher order (in $\varphi$) terms in the coupling function would have negligible contribution.

\section{Basic equations}\label{sect_3}
\subsection{The reduced system}
In the present paper we will be searching for static, spherically symmetric, asymptotically flat black holes.
The metric of a static, spherically symmetric space-time can be written in the form

\begin{equation}
ds^2 = g_{\mu\nu}dx^{\mu}dx^{\nu} = - f(r)e^{-2\delta(r)}dt^2 + {dr^2\over f(r) } +
r^2\left(d\theta^2 + \sin^2\theta d\phi^2 \right),
\end{equation}
where
\begin{equation}
f(r)=1-\frac{2 m(r)}{r}.
\end{equation}
$m(r)$ is the so-called local gravitational mass and should not be confused with the mass of the scalar field $m_{*}$.
We will study the magnetically charged black holes for which the electromagnetic field strength is given by
\begin{equation}
F = P \sin\theta d\theta \wedge d\phi
\end{equation}
and the magnetic charge is denoted by $P$.
The electrically charged solutions can be obtained through electric-magnetic duality rotations (we refer the reader to \cite{GR1, GR2, Bronnikov, SYT3} for more details on these duality transformations) of the type
\begin{equation}
\{ g_{\mu\nu},\, \varphi,\, F_{\mu\nu},\, P ,\,X,\, L(X)\}  \longleftrightarrow \
\{ g_{\mu\nu},\, \varphi,\, ~\star
G_{\mu\nu},\, \bar{Q} ,\,\bar{X},\, L(\bar{X})\}.\label{duality}
\end{equation}
The barred quantities are related to the dual system.
Here
\begin{equation}
G_{\mu\nu}=-2\frac{\partial \left[{\cal A}^{4}(\varphi)L\right]}{\partial F^{\mu\nu} },\label{constitution}
\end{equation}
\begin{equation}
\bar{X}=-\Bigl[\partial_{X}L(X)\Bigr]^2 X,
\end{equation}
and $\bar{Q}$ is the electric charge of the dual solution. The Born-Infeld lagrangian (\ref{LBI}) belongs to the class of lagrangians for which the system (\ref{F}) is invariant under the electric-magnetic duality transformations (\ref{duality}). As a consequence of that invariance the solutions for the metric functions and the scalar field in the electrically charged case and the magnetically charged case coincide.

The field equations reduce to the following coupled system of
ordinary differential equations
\begin{eqnarray}
&&f'' -2f\delta'' -3f'\delta'+
2f\delta'^2+\frac{2}{r}f'-\frac{4}{r}f\delta' =\nonumber\\
&&\,\,\,\,\,\,\,\,\,\,\,\,\,\,\,\,\,\,\,\,\,\,\,\,\,\,\,\,\,\,\,\,\,\,\,\,\,\,\,\,\,\,\,\,\,\,\,\,\,\,-4\left\{V(\varphi)+{\cal A}^{4}(\varphi) \left[2X\partial_{X}L(X)-L(X)\right]\right\}\label{EQ_tt},\\
&&f'' -2f\delta'' -3f'\delta'+2f\delta'^2+\frac{2}{r}f'=\nonumber\\
&&\,\,\,\,\,\,\,\,\,\,\,\,\,\,\,\,\,\,\,\,\,\,\,\,\,\,\,\,\,\,\,\,\,\,\,\,\,\,\,\,\,\,\,\,\,\,\,\,\,\,-4\left\{f\varphi\,'^{\,2}+V(\varphi)+{\cal A}^{4}(\varphi) \left[2X\partial_{X}L(X)-L(X)\right]\right\}\label{EQ_rr},\\
&&1-f-rf'+rf\delta'=2r^2\left[ V(\varphi)-
{\cal A (\varphi)}^{4}L(X)  \right] \label{EQ_thita},\\
&&\frac{d }{dr}\left( e^{-\delta}r^{2}f\frac{d\varphi }{dr} \right)=\frac{d V(\varphi)}{d\varphi}r^2+4 r^2 e^{-\delta} \alpha(\varphi)
{\cal A}^{4}(\varphi)\left[X\partial_{X}L(X)- L(X)\right]  \label{EQPhi_d}  ,
\end{eqnarray}
where $ X $ reduces to
\begin{equation}
X = {{\cal A}^{-4}(\varphi)\over 2} \frac{P^2}{r^4}.
\end{equation}\label{X}
It is a system of four equations for three unknown functions but, as it is well known, the self-consistency of the system is guaranteed by the Bianchi identity.
For numerical treatment of the problem, however, the following form is more convenient
\begin{eqnarray}
&&\frac{d\delta}{dr}=-r\left(\frac{d\varphi}{dr} \right)^2\label{EQDelta},\\
&&\frac{d m}{dr}=r^2\left[\frac{1}{2}f\left(\frac{d\varphi}{dr} \right)^2 + V(\varphi)-
{\cal A (\varphi)}^{4}L(X)  \right] \label{EQm},\\
&&\frac{d }{dr}\left( r^{2}f\frac{d\varphi }{dr} \right)=
r^{2}\left\{\frac{d V}{d\varphi}-4\alpha(\varphi){\cal A}^{4}(\varphi) \left[L -  X\partial_{X}L(X)\right] -
r f\left(\frac{d\varphi}{dr} \right)^3    \right\} \label{EQPhi}  .
\end{eqnarray}
In this form the last two equations are separated as an independent sub-system.

\subsection{Qualitative investigation}\label{subsect_qualit}

Some general properties of the solutions can be derived through an analytical investigation of the equations.
The major difference resulting from the different type of coupling between the scalar field and the electromagnetic field is as follows. In both theories, STT and dilaton gravity, the right-hand side of some of the generalized Einstein equations for the metric functions is proportional to the lagrangian of the electromagnetic field. The right-hand side of the equation for the scalar field in STT is proportional to the to the trace of the energy-momentum tensor of the electromagnetic field while in dilaton gravity it is proportional to the lagrangian of the electromagnetic field. The last fact allows in the case of dilaton gravity combinations between the field equations which are favorable for the analytical investigation of the properties of the solutions to be made.

Analytical assessments and results, though not very complete, are always welcome since the entire space of parameters cannot be examined numerically thoroughly. The examination of the properties of the solutions is additionally impeded by the presence of numerical instabilities for some values of the parameters. Such instabilities appear especially in the interior region and have been reported also in \cite{HH}.

For our assessments we will use the fact that for the Born-Infeld lagrangian of nonlinear electrodynamics (\ref{LBI}) the relations
\begin{equation}
 X\partial_{X}L(X)- L(X)>0\hspace{1cm}{\rm and}\hspace{1cm}
 2X\partial_{X}L(X)- L(X)<0\label{EDHAM1}
\end{equation}
hold.

For the analytical analysis we will also use the boundary conditions presented in the Section \ref{Num_sect} where the numerical treatment of the problem is presented. More specifically, we will want the scalar field to be vanishing at the spacial infinity.

It can be proved that the function $\varphi(r)$ is non-positive outside the event horizon. Let us first  exclude the possibility $\varphi(r)>0$ on the event horizon.
\begin{proposition}
$\varphi(r)$ cannot be positive on the event horizon.
\end{proposition}
\textit{\textbf{Proof:}}
In order to prove this we will use equation (\ref{EQPhi_d}).
Let us suppose that $\varphi(r)$ is positive on the event horizon and has one or more roots in the exterior region. Then we multiply equation (\ref{EQPhi_d}) by $\varphi(r)$ and integrate it in the interval $r\in[r_{H},r_{0}]$ where we denote
the radius of the event horizon by $r_{H}$ and the leftmost zero of $\varphi(r)$ outside the event horizon by $r_{0}$
\begin{eqnarray}
&&\int\limits_{r_{H}}^{r_{0}}\varphi\, \frac{d }{dr}\left( e^{-\delta}r^{2}f\frac{d\varphi }{dr}\right) dr \nonumber \\
&&\hspace{1cm}=\int\limits_{r_{H}}^{r_{0}}\left\{\varphi \, \frac{d V(\varphi)}{d\varphi} r^2+4 r^2  e^{-\delta}\varphi\, \alpha(\varphi)
{\cal A}^{4}(\varphi)\left[X\partial_{X}L(X)- L(X)\right]\right\}dr. \label{Int_ots1}
\end{eqnarray}
After integrating by parts we get
\begin{eqnarray}
&&\left. \left( \varphi\, e^{-\delta}r^{2}f\frac{d\varphi }{dr} \right) \right|_{r_{0}}-
\left.\left(\varphi\, e^{-\delta}r^{2}f\frac{d\varphi }{dr} \right) \right|_{r_{H}}-
\int\limits_{r_{H}}^{r_{0}} \left[e^{-\delta}r^{2}f\left(\frac{d\varphi }{dr}\right)^{2}\right] dr \nonumber \\
&&\hspace{1cm}\hspace{1cm}\hspace{1cm}=-\int\limits_{r_{H}}^{r_{0}} \left[e^{-\delta}r^{2}f\left(\frac{d\varphi }{dr}\right)^{2}\right] dr\nonumber \\
&&\hspace{1cm}=\int\limits_{r_{H}}^{r_{0}}\left\{\varphi\,\frac{d V(\varphi)}{d\varphi} r^2+4 r^2 e^{-\delta}\varphi\, \alpha(\varphi)
{\cal A}^{4}(\varphi)\left[X\partial_{X}L(X)- L(X)\right]\right\} dr.\label{Int_ots2}
\end{eqnarray}

Taking into account (\ref{condPOT}) and (\ref{EDHAM1}) and the fact that according to the admission $\varphi$ is positive in  $r\in[r_{H},r_{0}]$ we see that the sign of the right-hand side (RHS) of (\ref{Int_ots2}) is positive. The left-hand side (LHS), however, is negative since $f(r)>0$ outside the event horizon. The contradiction we reach means that our admission is incorrect. If we admit that $\varphi(r)$ is positive on the event horizon and has no roots following the same procedure we will again reach a contradiction. In order to see this it is enough to let $r_{0}\rightarrow \infty$ in (\ref{Int_ots1}) and to impose the boundary condition $\varphi(r)\rightarrow 0$ at infinity.
$\square$

\begin{proposition}
Function $\varphi(r)$ is negative and has no zeros in the exterior region.
\end{proposition}
\textit{\textbf{Proof:}}
Let us admit that $\varphi(r)$ has zeros.  We know already that $\varphi(r)$ is negative on the event horizon and also require it to satisfy the vanishing boundary conditions at infinity. Then one of the following two situations must be realized. Either the $\varphi(r)$ has at least one positive maximum or it has at least one zero maximum. First, let us exclude the possibility for existence of positive maxima of $\varphi$. Again, we can use (\ref{Int_ots1}) but this time let us integrate in the interval $r\in[r_{0},r_{e}]$. Quantity $r_{e}$ denotes the point in which $\varphi(r)$ has a positive maximum, i.e., $\varphi'(r_{e})=0$, where $(\cdot)'$ denotes the derivative with respect to the radial coordinate $r$. Here $r_{0}$ denotes the first zero of $\varphi$ to the left of $r_{e}$.
After integrating by parts we get
\begin{eqnarray}
&&\hspace{1cm}\hspace{1cm}\hspace{1cm}-\int\limits_{r_{0}}^{r_{e}} \left[e^{-\delta}r^{2}f\left(\frac{d\varphi }{dr}\right)^{2}\right] dr\nonumber \\
&&\hspace{1cm}=\int\limits_{r_{0}}^{r_{e}}\left\{\varphi\,\frac{d V(\varphi)}{d\varphi} r^2+4 r^2 e^{-\delta}\varphi\, \alpha(\varphi)
{\cal A}^{4}(\varphi)\left[X\partial_{X}L(X)- L(X)\right]\right\} dr.\label{Int_ots3}
\end{eqnarray}
Since $\varphi(r)$ is positive in  $r\in[r_{0},r_{e}]$ the sign of the RHS of (\ref{Int_ots3}) is positive. The LHS, however, is negative. Again we reach a contradiction which means that the admission for presence of a positive maximum of $\varphi$ is false.

It remains only to exclude the case in which $\varphi$ becomes zero at a maximum, which we will again denote as $r_e$. At a zero maximum $\varphi(r_e)=0$, $\varphi'(r_e)=0$ and $\varphi''(r_e)<0$,  so from equation (\ref{EQPhi_d}) taking into account also (\ref{condPOT}) we obtain
\begin{eqnarray}
\left. f\varphi'' \right|_{r_e}=\left.4  \alpha(\varphi)
{\cal A}^{4}(\varphi)\left[X\partial_{X}L(X)- L(X)\right]\right|_{r_e}.
\end{eqnarray}
The LHS of the above expression is negative while the RHS is positive. This contradiction allow us to exclude the cases of zeros at maxima of $\varphi$. This completes the proof of the proposition.
$\square$

In our numerical calculations the following restriction in the behavior of $f$ will be useful.
\begin{proposition}
The metric function $f(r)$ cannot have extrema with $f(r)>1$.
\end{proposition}

 Let us consider equation (\ref{EQ_thita}). Using (\ref{EQDelta})  we can write it in the form
\begin{equation}
 1-f-rf'-r^2f\varphi'^2=2r^2\left[ V(\varphi)-{\cal A (\varphi)}^{4}L(X)  \right].
\end{equation}
From here we obtain
\begin{equation}
 -rf'=(f-1)+2r^2\left[ {1\over 2}f\varphi'^2+V(\varphi)-{\cal A (\varphi)}^{4}L(X)  \right].\label{extremum_f}
\end{equation}
When $f(r)>1$ the RHS of (\ref{extremum_f}) is positive so the derivative of $f(r)$ is strictly negative,  $f'(r)<0$. Hence, $f(r)$ cannot have extrema with $f(r)>1$.$\square$

Once becoming greater than one, $f(r)$ must continue to rise, no matter whether we go forward or backward in $r$. The inward integration can stop once the point $f(r)=1$ is surpassed.

\begin{proposition}
From the equations we can see that the function $\delta(r)$ is smooth non-increasing and has inflexion points where $\varphi(r)$ has extrema.
\end{proposition}
\textit{\textbf{Proof:}}
From equation (\ref{EQDelta}) we see that its first derivative is negative so the first part of the proposition is easily proven.
Now let us differentiate equation (\ref{EQDelta})
\begin{equation}
\frac{d^2\delta}{dr^2}=-\left(\frac{d\varphi}{dr} \right)^2-2\, r \frac{d\varphi}{dr}\cdot \frac{d^2\varphi}{dr^2}.\label{EQDelta_2}
\end{equation}
From (\ref{EQDelta}) and (\ref{EQDelta_2}) we see that in the extrema of $\varphi(r)$ (the points in which $\varphi'(r)=0$)
$\delta'(r)=\delta''(r)=0$, i.e., $\delta(r)$ has an inflexion point.
$\square$

\subsection{Asymptotic solutions}\label{Asyptotic}
The presence of potential changes considerably the asymptotic behavior of the scalar field in comparison to the massless scalar field case. In the asymptotic region $r\rightarrow\infty$ we find the following behavior for the functions
\begin{equation}
f(r)=1-{2M \over r}+{P^2 \over r^2}+{P^4 \over 40 \, b\, r^6}+\mathcal{O}(r^{-14}),  \label{asymp_f}\\
\end{equation}
\begin{equation}
\delta(r)=\mathcal{O}(r^{-16}),   \label{asymp_delta} \\
\end{equation}
\begin{equation}
\varphi(r)=-{\alpha(0)\, P^4 \over 4 m_{*}^2\,b\,r^8}+\mathcal{O}(r^{-9}). \label{asymp_phi}
\end{equation}

These asymptotic solutions are later used as initial approximations for the numerical integration.

\subsection{Units}\label{Asyptotic}
In the system of units that we work $G_*=c={\mu_0\over4\pi}=1$, where $\mu_0$ is the magnetic constant. In this system
$$
[P]=m; \,\,\,\,[M]=m;\,\,\,\,[m_*]=m^{-1};\,\,\,\,[b]=m^{-2}.
$$

In order to obtain dimensionless quantities we can use the fact that the differential equations system (\ref{EQDelta})-(\ref{EQPhi}) is invariant under the rigid re-scaling
$r\to \lambda r$, $m \to \lambda m$, $P \to \lambda P$ and $b \to \lambda^{-2} b$ where $0<\lambda< \infty$. Therefore, given
a solution to (\ref{EQDelta})-(\ref{EQPhi}) with one set of physical parameters $(r_{h}, M, P, m_*, b, T)$, the rigid re-scaling produces  new solutions with parameters $(\lambda r_{h}, \lambda M, \lambda P, \lambda^{-1} m_* ,\lambda^{-2} b, \lambda ^{-1}T)$. Here $T$ denotes the temperature of the horizon.
The dimensionless parameters can be obtained through such re-scaling if $[\lambda]=m^{-1}$. Here, we introduce scale by choosing $\lambda=\sqrt{b}$. We will keep the notation for the dimensionless quantities unchanged.

\section{Extremal solutions}\label{Extremalni}

The massive-scalar-field black-hole solution has a much richer causal structure compared to the case with massless scalar field, since
in the former case the presence of inner and extremal horizons is allowed. For a two-fold degenerate horizon from equations (\ref{EQm}) and (\ref{EQPhi})
having in mind that on the extremal horizon
\begin{equation}
f(r_{e})=0 \,\, ,  \,\,\,
{dm(r_{e})\over dr}=\frac{1}{2}\,\,
\end{equation}
hold, we obtain
\begin{eqnarray}
&&r_{e}^2\left[V(\varphi_{e})- {\cal A}^{4} (\varphi_{e})L(X_{e})  \right]=\frac{1}{2}\label{extr1}\,\, ,\\
&&r_{e}^{2}\left\{\frac{dV(\varphi_{e})}{d\varphi}-
4\alpha(\varphi_{e}){\cal A}^{4}(\varphi_{e}) \left[L(X_{e})-X_{e}\partial_{X}L(X_{e})\right]\right\}=0. \label{extr2}
\end{eqnarray}
Here and forth, unlike the previous section, $r_{e}$ will denote the extremal horizon. Again, subscript ``$e$'' denotes the value of the functions evaluated at $r_{e}$. Equations (\ref{extr1})-(\ref{extr2}) can be solved with respect to $r^2_{e}$
\begin{eqnarray}
r_{e}^2= \frac{-F_{1}\pm \sqrt{F_{1}^2-4\alpha(\varphi_{e}) F_{2}} }{ 2F_{2}}, \label{Re}
\end{eqnarray}
where
\begin{eqnarray}
&&F_{1}= \left[\displaystyle\frac{d V (\varphi_{e})}{d\varphi}-4\alpha(\varphi_{e}) V(\varphi_{e}) \right],\\
&&F_{2}= {\left[4 b \displaystyle\frac{d V (\varphi_{e})}{d\varphi}{\cal A}^{4}(\varphi_{e})-
2V(\varphi_{e}) \displaystyle\frac{d V(\varphi_{e})}{d\varphi}
+4 \alpha(\varphi_{e}) V(\varphi_{e})^2 \right]}.
\end{eqnarray}\label{extrem_conds}
The root for which $0<r_{e}^2<\infty$ for all values of $\varphi_{e}$ should be chosen.
Then we substitute it
in equation (\ref{extr1}) and obtain the following
sophisticated non-linear equation for $\varphi_{e}$
\begin{multline}\label{eqphiextrem}
1= \frac{-F_{1} + \sqrt{F_{1}^2-4\alpha(\varphi_{e}) F_{2}} }{ 2F_{2}}\times\\
\left\{ 2V(\varphi_{e})-4b {\cal A}^{4}(\varphi_{e})
\left[ 1-
\sqrt{1+\displaystyle\frac{P^2}{2b}{\cal A}^{4}(\varphi_{e})\left( \frac{2F_{2}} {-F_{1} +
\sqrt{F_{1}^2-4\alpha (\varphi_{e})F_{2}}  }\right)^{2}} \right]\right\},
\end{multline}
which we treat numerically for the STT with $\alpha(\varphi)=\mathrm{const}\equiv\alpha$.
The solution for $\varphi_{e}$ is substituted back in (\ref{Re}) which, on its turn,
gives us the radius of the degenerate horizons. Equation (\ref{eqphiextrem}) may have two roots for $\varphi_{e}$ which give two roots for $r_{e}^2$ when substituted back in (\ref{Re}). Which of them describes a degenerate event horizon? Using equation (\ref{EQ_tt}) we can find the value $f_{e}''$
\begin{equation}
f_{e}'' =-4\left\{V(\varphi_{e})+{\cal A}^{4}(\varphi_{e}) \left[2X_{e}\partial_{X}L(X_{e})-L(X_{e})\right]\right\}.
\end{equation}
Hence, we can use the fact that at the degenerate event horizon $f_{e}''>0$ to obtain a criterium which tells us whether the roots for $\varphi_{e}$ and $r_{e}$ obtained from (\ref{eqphiextrem}) and (\ref{Re}) correspond to a degenerate event horizon. The roots for which $f_{e}''<0$ could correspond to an internal degenerate horizon resulting from the merger of two internal horizons if they exist at all.

\begin{figure}[htbp]%
\vbox{ \hfil \scalebox{0.5}{ {\includegraphics{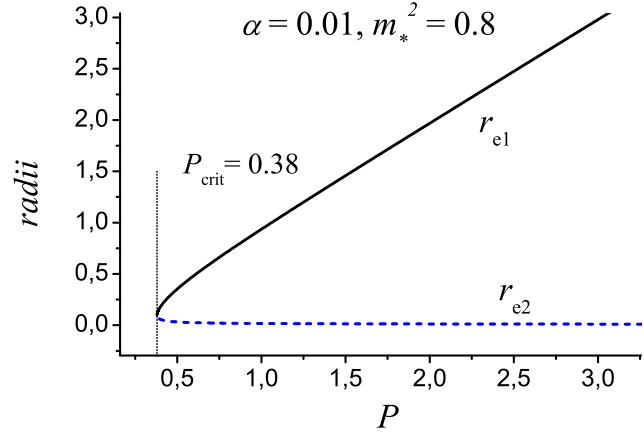}} }\hfil}%
\bigskip%
\caption{%
Solutions for the positions of the degenerate horizons as a function of the magnetic charge $P$ of the black hole. }
\label{r_extr_P}%
\end{figure}

Equation (\ref{eqphiextrem}) for $\varphi_{e}$ has two roots. The
corresponding values $r_{e_{1}}$ and $r_{e_{2}}$ are given in figure (\ref{r_extr_P}). It should be noted that the analysis for existence of degenerate horizons does not take in consideration the boundary conditions.
In other words, it is possible that for some values of the parameters the extremal black-hole solutions to be
asymptotically non-flat.

From the graphics we see that for fixed values of $\alpha$ and $m_{*}$, a critical value of the magnetic charge $P_{\rm{crit}}$ exists,
such that for $P>P_{\rm{crit}}$ solutions with degenerate event horizons exist and for $P<P_{\rm{crit}}$ solutions with
degenerate event horizons do not exist. For example, when $\alpha=0.01$ and $ m_{*}^2=0.8$, $P_{\rm{crit}}\approx0.38$. A degenerate horizon is usually formed when for some values of the black-hole mass $M$ two regular horizons
merge. So, it is likely to expect that for $P>P_{\rm{crit}}$ black holes with two regular horizons exist for some values of $M$. On the contrary,
when $P<P_{\rm{crit}}$ we should not expect the existence of solution with more than one horizon. Our further numerical investigation
confirms the expectations for the case $P>P_{\rm{crit}}$. We should also note that for fixed $P>P_{\rm{crit}}$ with the decrease of $M$ an extremal solution is reached while for $P<P_{\rm{crit}}$ --- a naked singularity.

\begin{figure}[htbp]%
\vbox{ \hfil \scalebox{0.5}{ {\includegraphics{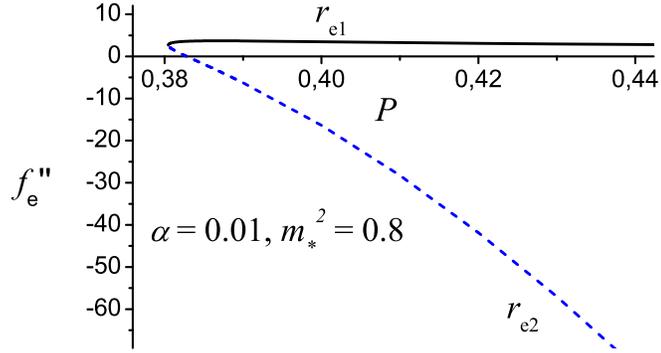}} }\hfil}%
\bigskip%
\caption{%
Derivative $f''_e$ evaluated on $r_{e1}$ and $r_{e2}$ as a function of the magnetic charge $P$ of the black hole. On a degenerate event horizon $f''_e>0$ must hold.}
\label{f_second_extr_P}%
\end{figure}

The $P-f_{e}''$ dependence is presented in figure (\ref{f_second_extr_P}). As we can see, for the whole interval of its existence the root
$r_{e_{1}}$ cannot be excluded as a candidate for a degenerate event horizon. The situation with the second root is more complicated.
For most values of $P$, $r_{e_{2}}$ cannot correspond to a degenerate event horizon. For a small interval close to $P_{\rm{crit}}$, however,
$f_e''$ is positive. There are several ways to interpret $r_{e_{2}}$ in that interval. It could correspond to a degenerate internal horizon where the function $f$ has a minimum. Since this analysis does not take into account the boundary conditions, another possibility would be that this root  corresponds to a degenerate event horizon of a black hole with a different asymptotic behavior.  The point where
$f_e''$ turns to zero could correspond to a triply degenerate event horizon. Unfortunately we have not been able to check any of these speculations
since the numerical simulations fail in that region of the parameter space.


\begin{figure}[htbp]%
\vbox{ \hfil \scalebox{0.5}{ {\includegraphics{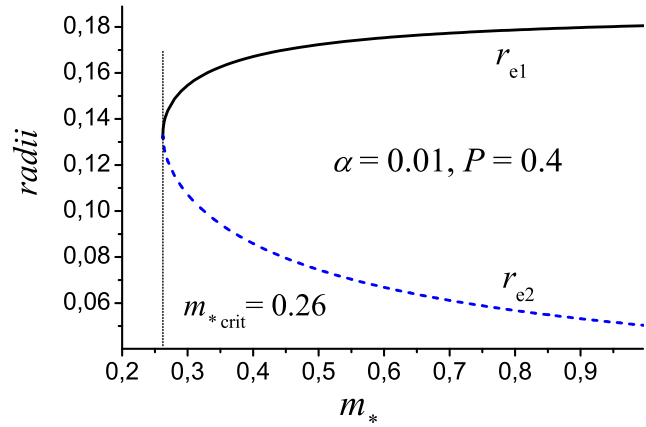}} }\hfil}%
\bigskip%
\caption{%
Solutions for the positions of the degenerate horizons as a function of the scalar field mass $m_{*}$.
} \label{r_extr_m_star}%
\end{figure}

Alternatively, we could keep the values of $\alpha$ and $P$ fixed and study the presence of degenerate event horizons for different $m_{*}$. The numerical solutions for that case are shown in figure (\ref{r_extr_m_star}).  Again, a critical value for scalar-field mass appears $m_{*,\,\rm{crit}}$, such that for $m_{*} > m_{*,\,\rm{crit}}$ solutions with degenerate event horizons exist and for $m_{*} < m_{*,\,\rm{crit}}$ no solutions with degenerate event horizons exist. For example, when $\alpha=0.01$ and $P=0.4$, $m_{*,\,\rm{crit}}\approx0.26$. Critical mass $m_{*,\,\rm{crit}}$ decreases with the increase of the magnetic charge $P$.

\begin{figure}[htbp]%
\vbox{ \hfil \scalebox{0.5}{ {\includegraphics{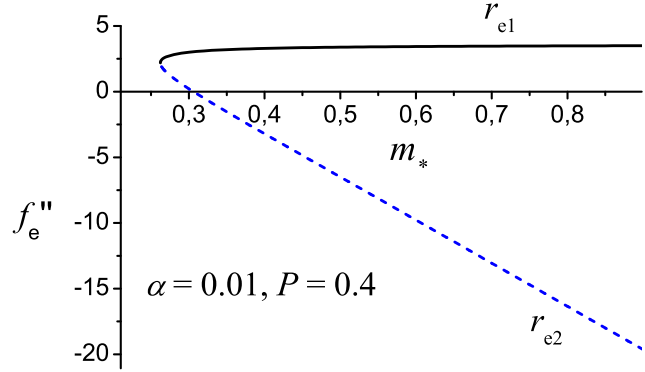}} }\hfil}%
\bigskip%
\caption{Derivative $f''_e$ evaluated on $r_{e1}$ and $r_{e2}$ as a function the scalar field mass $m_{*}$. On a degenerate event horizon
$f''_e>0$ must hold. }
\label{f_second_extr_m_star}%
\end{figure}

The $m_{*}-f_{e}''$ dependence is presented in figure (\ref{f_second_extr_m_star}). The comments about the two roots are the same as in the
case presented in figure (\ref{f_second_extr_P}).

The critical value $P_{\rm{crit}}$ depends on $m_*$ and $\alpha$. In order to determine this dependence we will use again equations (\ref{extr1}) and (\ref{extr2}). Let us introduce a new variable
\begin{equation}
k= \sqrt{1+ \frac{X_e}{b}}\label{k}.
\end{equation}
The variable $k$ can be expressed as a function of $\varphi_e$ from (\ref{extr2}). Using (\ref{X}) and (\ref{k}) we can express $r_e^2$ in the following way
\begin{equation}
r_e^2 = {{\cal A}^{-2}(\varphi_e) |P| \over \sqrt{2} \sqrt{k^2-1}}.\label{re2}
\end{equation}

\begin{figure}[htbp]%
\vbox{ \hfil \scalebox{0.5}{ {\includegraphics{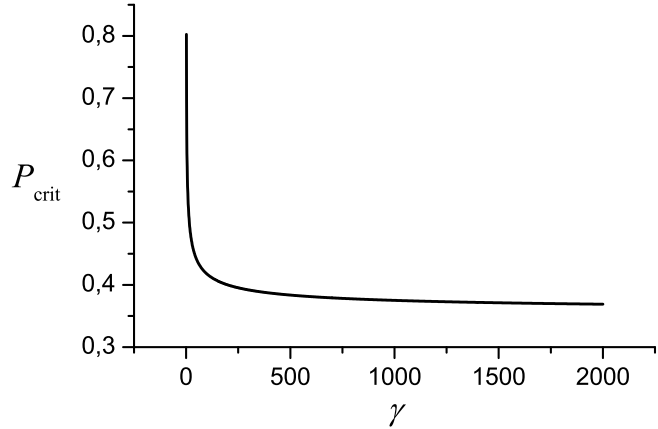}} }\hfil}%
\bigskip%
\caption{The $\gamma - P_{\rm{crit}}$ critical curve }
\label{crit_curve}%
\end{figure}

Then, in (\ref{extr1}) we express $L(X_e)$ with $k$ and $r_e^2$ from (\ref{re2}). Introducing a new variable $y=\alpha\varphi_e$ we obtain
\begin{equation}
{1\over 2} = \sqrt{2}\,|P|\,{\gamma y^2 e^{-2y}+e^{2y}(k-1)\over \sqrt{k^2-1}}=\sqrt{2}\,|P|\, F(y, \gamma),\label{criteq}
\end{equation}
where
$$
\gamma={m_*^2\over 4 \alpha^2}.
$$
In (\ref{criteq}) there are two free parameters $|P|$ and $\gamma$. Let us fix the value of $\gamma$. Then $F(y, \gamma)$ has one maximum ($y<0$).
Depending on $|P|$ the RHS of (\ref{criteq}) can be greater, equal to or less than the LHS and the equations have,
respectively, two, one, or no solutions. So  $P_{\rm{crit}}$\footnote{$P$ is non-negative.} is the value for which (\ref{criteq})
has only one solution. The $\gamma - P_{\rm{crit}}$ critical curve is plotted in figure (\ref{crit_curve}).

\section{Numerical integration}\label{Num_sect}

The nonlinear system (\ref{EQDelta})-(\ref{EQPhi}) is
inextricably coupled. Our aim is to obtain numerical solutions that describe asymptotically flat black holes. We split the problem in two boundary sub-problems -- in the exterior and in the interior region of the black hole, which we solve subsequently.

\subsection{External problem}
First, the exterior region. For that region we can formulate a boundary value problem (BVP) for (\ref{EQDelta})-(\ref{EQPhi}) with the boundary conditions
$$\lim_{r \to \infty}m(r) =M \quad (M \>{\rm is\> the\> mass\> of\> the\> black\> hole\> in\> the\> Einstein\> frame}),$$
$$ \lim_{r \to \infty}\delta(r)=\lim_{r \to \infty}\varphi(r)=0.$$
for the right-hand boundary at spatial infinity and
$$f(r_H)=0$$
on the horizon.
The left-hand boundary, namely the event horizon, is {\it a priori} unknown. Such BVPs are known in mathematical physics as BVPs of Stefan kind (see, for example \cite{tihsam}). For the location of the event horizon an additional condition is needed. It can come from the requirement all functions to be regular on the event horizon
$$\left.\left(\frac{df}{dr}\!\cdot\! \frac{d \varphi}{d r}\right)\right|_{r=r_H} =\left.\frac{d V(\varphi)}{d\varphi}\right|_{r=r_H}+\left.\left\{ 4 \alpha(\varphi) {\cal A}^4(\varphi) [X \partial_X L(X)-L(X) ]\right\}\right|_{r=r_H}.$$
We can also think of the so formulated BVP as a non-linear analogue of a spectral problem with regard to parameter $r_H$.

Having in mind these features of the above posed BVP we treat it by using the Continuous Analog of
Newton Method (see, for example \cite{gavurin},\cite{jidkov},\cite{YFBT}). After an appropriate linearization the original BVP
is rendered to solving standard vector two-point BVPs. On a discrete level almost diagonal linear algebraic
systems with regard to increments of sought functions $\delta(r)$, $m(r)$, and $\varphi(r)$ have to be inverted.

\subsection{Internal problem}
The exterior solutions can be continued inwards. The values of the functions and their derivatives on the event horizon are obtained \textit{a priori} in the exterior problem. This allows an Initial-Value Problem (IVP) for the same system (\ref{EQDelta})-(\ref{EQPhi}) to be formulated in the interior region $r<r_H$ \textit{a posteriori}. The event horizon, however, is a singular point for the equation of the scalar field (\ref{EQPhi}) since the coefficient in front of $\varphi''$ turns to zero there ($f(r_H)\equiv0$) and the equation loses its leading order term. So, to pose a regular IVP we shift the initial point $r_H$ by small enough $\varepsilon>0$ and choose for initial point $r_H-\varepsilon$ instead $r_H$. On the other hand the functions in question are smooth  in the interval $(r_H-\varepsilon,r_H)$ and hence the following series expansions hold
\begin{eqnarray}
&& m(r_{H}-\varepsilon)= m(r_{H})-m'(r_{H})\varepsilon+o(\varepsilon^2),\\
&&\delta(r_{H}-\varepsilon)= \delta(r_{H})-\delta'(r_{H})\varepsilon+o(\varepsilon^2),\\
&&\varphi(r_{H}-\varepsilon)= \varphi(r_{H})-\varphi'(r_{H})\varepsilon +o(\varepsilon^2),\\
&&\Phi(r_{H}-\varepsilon)= \Phi(r_{H})-\Phi'(r_{H})\varepsilon+o(\varepsilon^2),\quad{\rm where}\quad \Phi(r)=\varphi'(r).
\end{eqnarray}
A similar shift is made also at the every inner horizon (if such is reached). The latter admits an algorithmic sequence of IVPs for finding possible inner horizons. For the numerical treating of the above posed IVP again the Continuous Analog of Newton Method is used.


\subsection{Some results}
Let us first consider the case $P>P_{\rm{crit}}$. The numerical investigation of the solutions shows that for fixed values of the parameters $\alpha$ and $m_{*}$ the general structure of the solutions depends strongly on the charge-to-mass ratio $P/M$. For low enough $P/M$ the obtained black-hole solutions have a single horizon, namely the event horizon. For high enough $P/M$ (for $P/M$ close to $1$) the black holes have two horizons or one degenerate horizon (an extremal solution is reached when we decrease the mass $M$ and keep $P$ fixed).

\begin{figure}[htbp]%
\vbox{ \hfil \scalebox{1.2}{ {\includegraphics{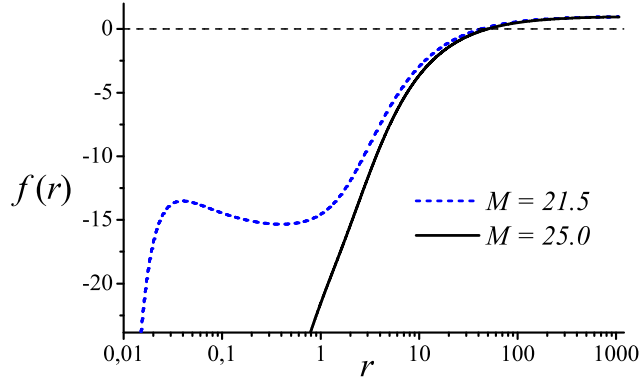}} }\hfil}%
\bigskip%
\caption{%
The metric function $f$ for $\alpha=0.01,  m_{*}^2=0.8, P=6.0$ and two different masses of the black hole $M=21.5$ and $M=25.0$. The abscissa is in logarithmic scale.
} \label{one_hor_f}%
\end{figure}

\begin{figure}[htbp]%
\vbox{ \hfil \scalebox{1.2}{ {\includegraphics{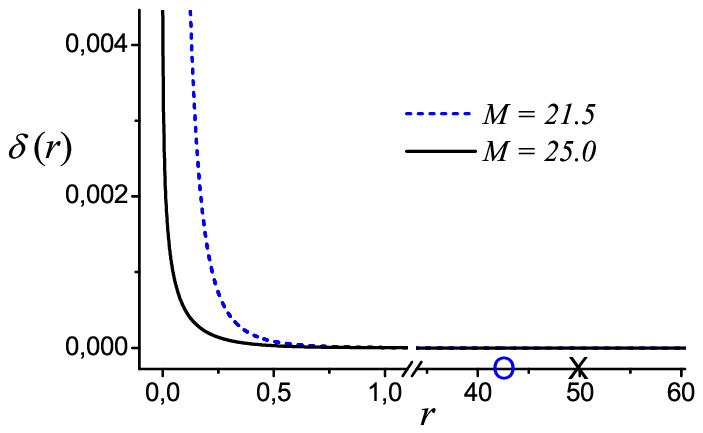}} }\hfil}%
\bigskip%
\caption{%
The metric function $\delta$ for $\alpha=0.01,  m_{*}^2=0.8, P=6.0$ and two different masses of the black hole $M=21.5$ and $M=25.0$.
} \label{one_hor_delta}%
\end{figure}

\begin{figure}[htbp]%
\vbox{ \hfil \scalebox{1.2}{ {\includegraphics{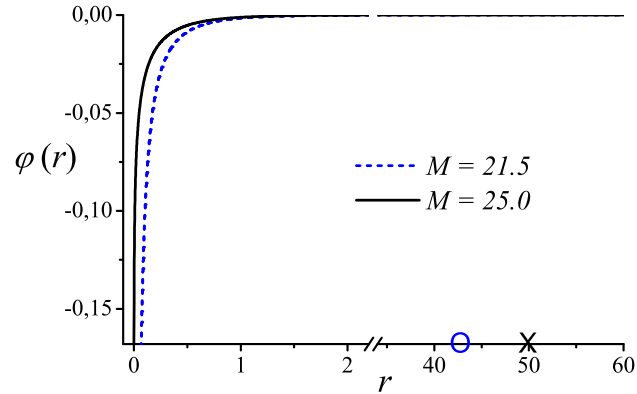}} }\hfil}%
\bigskip%
\caption{%
The scalar field $\varphi$ for $\alpha=0.01,  m_{*}^2=0.8, P=6.0$ and two different masses of the black hole $M=21.5$ and $M=25.0$.
} \label{one_hor_phi}%
\end{figure}

Two examples of solutions with a single horizon are shown (the functions $f$, $\delta$ and $\varphi$, respectively) in figures (\ref{one_hor_f}), (\ref{one_hor_delta}) and (\ref{one_hor_phi}). The values of the parameters in the presented examples are $\alpha=0.01, m_{*}^2=0.8, P=6.0$ and two different masses of the black hole $M=21.5$ and $M=25.0$. As it can be seen in figure (\ref{one_hor_f}) for $M=21.5$ the function $f$ has negative extrema below the event horizon (the abscissa in that figure is in logarithmic scale). In figures (\ref{one_hor_delta}) and (\ref{one_hor_phi}) the radius of the solution with $M=21.5$ is designated with ``$\bigcirc$'' and for $M=25.0$ -- with ``$\mathbf{\times}$.''

\begin{figure}[htbp]%
\vbox{ \hfil \scalebox{1.2}{ {\includegraphics{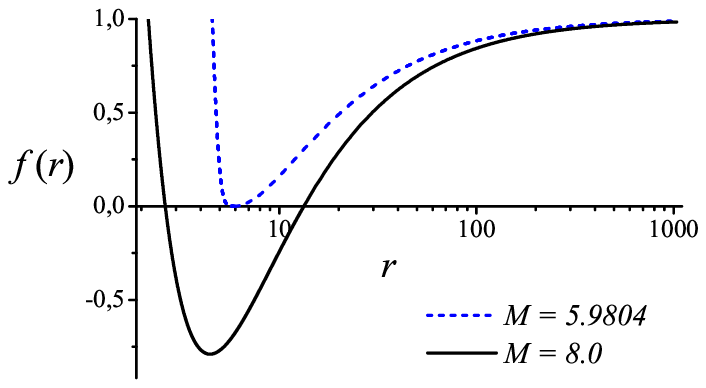}} }\hfil}%
\bigskip%
\caption{%
The metric function $f$ for $\alpha=0.01,  m_{*}^2=0.8, P=6.0$ and two different masses of the black hole $M=5.9804$ and $M=8.0$.
The abscissa is in logarithmic scale.
} \label{two_hor_f}%
\end{figure}

\begin{figure}[htbp]%
\vbox{ \hfil \scalebox{1.2}{ {\includegraphics{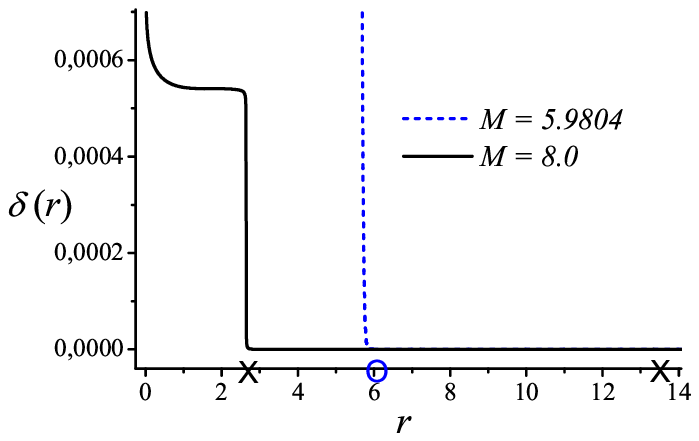}} }\hfil}%
\bigskip%
\caption{%
The metric function $\delta$ for $\alpha=0.01,  m_{*}^2=0.8, P=6.0$ and two different masses of the black hole $M=5.9804$ and $M=8.0$.
} \label{two_hor_delta}%
\end{figure}

\begin{figure}[htbp]%
\vbox{ \hfil \scalebox{1.2}{ {\includegraphics{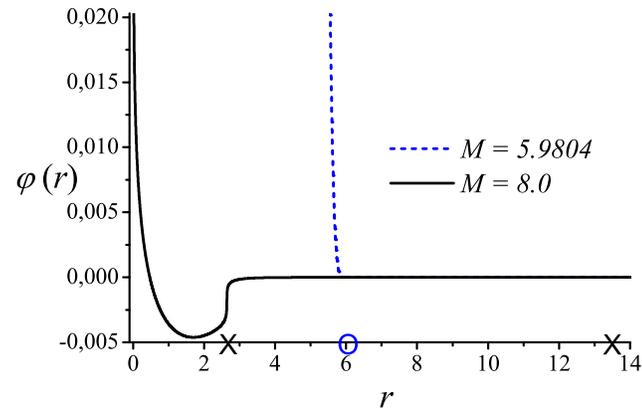}} }\hfil}%
\bigskip%
\caption{%
The scalar field  $\varphi$ for $\alpha=0.01,  m_{*}^2=0.8, P=6.0$ and two different masses of the black hole $M=5.9804$ and $M=8.0$.
}\label{two_hor_phi}%
\end{figure}

Solutions with two horizons and with one degenerate horizon are given in figures (\ref{two_hor_f}), (\ref{two_hor_delta}) and (\ref{two_hor_phi}). The values of the parameters in these cases are  $\alpha=0.01,  m_{*}^2=0.8, P=6.0$ and black-hole masses $M=5.9804$ for the extremal black hole and $M=8.0$ for the black hole with two horizons. We can see that in accordance with our observations from Section \ref{sect_3} $\delta(r)$ has an inflexion point in the extremum of $\varphi(r)$. In figures (\ref{two_hor_delta}) and (\ref{two_hor_phi}) the radius of the extremal black hole is designated with ``$\bigcirc$'' while the two radii of the solution with $M=8.0$ are designated with ``$\mathbf{\times}$''-es.

The presence of numerical instabilities did not allow us to study thoroughly the interior region for the case $P<P_{\rm{crit}}$.


\section{Thermodynamics}

\begin{figure}[htbp]%
\vbox{ \hfil \scalebox{1.2}{ {\includegraphics{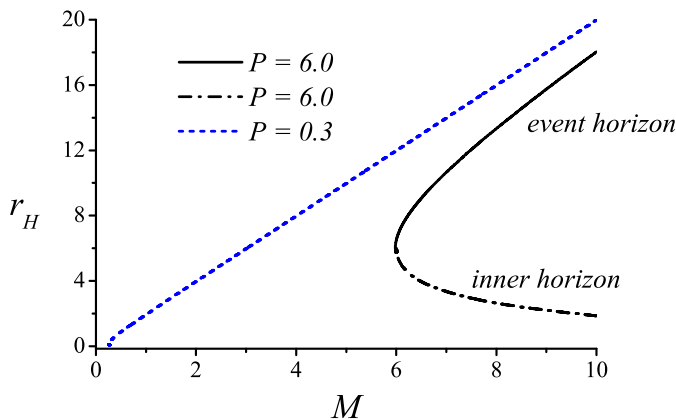}} }\hfil}%
\bigskip%
\caption{%
The radii of the horizons of black holes as function of the mass $M$. For $P=6$ the black hole has two horizon -- event horizon and inner horizon. For $P=0.3$ the black holes have a single horizon -- the event horizon.
} \label{rm}%
\end{figure}

\begin{figure}[htbp]%
\vbox{ \hfil \scalebox{1.2}{ {\includegraphics{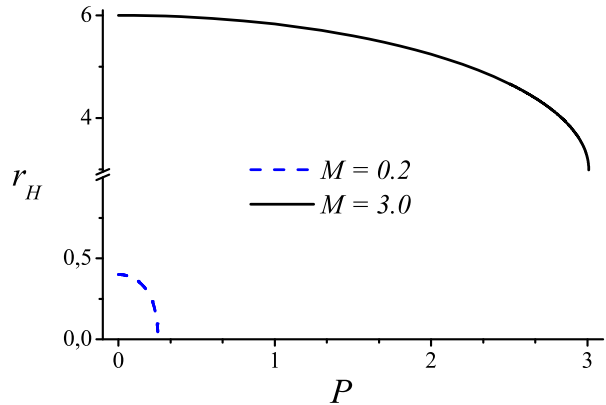}} }\hfil}%
\bigskip%
\caption{%
The $r_{H} - P$ relation.
} \label{rp}%
\end{figure}

Since the entropy of the black hole is related to the area, and respectively to the radius, of the event horizon in figures (\ref{rm}) and (\ref{rp}) we have given the $M-r_{H}$ and the $P-r_{H}$ diagrams. In the presented cases $\alpha=0.01, m_{*}^2=0.8$. For these values of the parameters $P_{\rm{crit}}\approx 0.38$. We have given two graphics -- one for $P=6.0>P_{\rm{crit}}$ and one for $P=0.3<P_{\rm{crit}}$. In the former case for masses in the interval $M\in [5.9804, 21.1]$ the black holes have two regular horizons (in figure (\ref{rm}) they are designated as \emph{event horizon} and \emph{inner horizon}). The two horizons merge and an extremal black hole occurs at $M=5.9804$. The value of event horizon obtained through numerical integration of the equations and the one obtained through solution of the algebraic problem in Section \ref{Extremalni} coincide within the approximation and round-off error, which is a very good test for the correctness of the obtained numerical results. In the latter case the black holes have a singe non-degenerate horizon.

Let us try to obtain some intuition of the general behavior of our solutions through comparison to the solution of Reissner-Nordstr\"{o}m (RN).
In RN black holes exist for $M\geq P$. An extremal black hole occurs at $P/M=1$. From the $M-r_{H}$ relation (\ref{rm}) we see that,
similarly, in our case black-hole solutions exist for $M>P$. Unlike RN, however, in our case black holes exist also for $M$ a bit less than $P$ but still the extremal black hole occurs at $P/M$ close to $1$. So if we fix the value of $M$
the maximal value of $P$ is of the same order. 
We can define also
$M_{\rm{crit}}$ as the mass of the extremal black hole when $P=P_{\rm{crit}}$. The relation $P_{\rm{crit}}/M_{\rm{crit}}\approx1$ also holds. This
dependence can be seen also in figure (\ref{rp})  where the $P-r_{H}$ relation is
shown \footnote{On that figure only the radii of the event horizons are shown.}. Let us set $M=0.2$. Then according to our admission $P$
cannot be much greater than $0.2$ so it will be less than $P_{\rm{crit}}$. Indeed, from figure (\ref{rp}) we see that in that case a naked
is reached when $P$ approaches its maximal value (see the comments at the end of Section \ref{Extremalni}). For $M=3.0$ an extremal black
hole appears when $P$ approaches its maximal value.

A knowledge about the general stability of the obtained solutions can be obtained through the application of the so-called ``turning point" method (we refer the reader to \cite{katzI, Sorkin:1981jc, Sorkin:1982ut, Arcioni} for a detailed discussion on the ``turning point" method and also to \cite{katzII, Tamaki, SYT3} for the application of the method to study the thermodynamical stability of black holes). According to that method, in micro-canonical ensemble \footnote{We keep the magnetic charge $P$ of the black hole fixed.}, a change of the stability reveals itself on the $M - T^{-1}$ diagram as bifurcation or turning points. Here ``bifurcation point'' is used to denote a point where brunching of equilibrium sequences occurs while a ``turning point'' is such point where two equilibrium sequences merge with a vertical tangent. The absence of such points on the $M - T^{-1}$ diagram means that if at least one point on the equilibrium sequence is stable then the whole equilibrium sequence is stable.

\begin{figure}[htbp]%
\vbox{ \hfil \scalebox{1.2}{ {\includegraphics{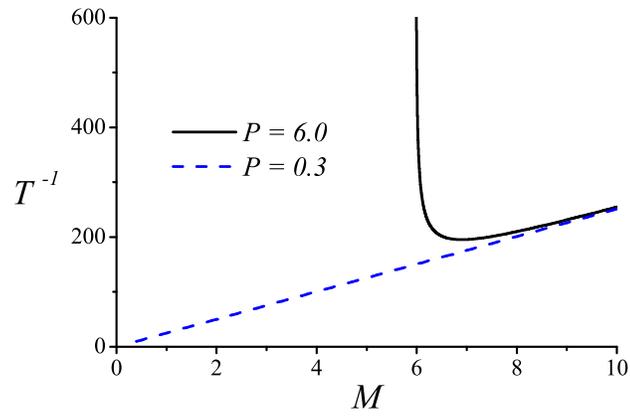}} }\hfil}%
\bigskip%
\caption{%
The $M - T^{-1}$ relation.
} \label{MinvT}%
\end{figure}

\begin{figure}[htbp]%
\vbox{ \hfil \scalebox{1.2}{ {\includegraphics{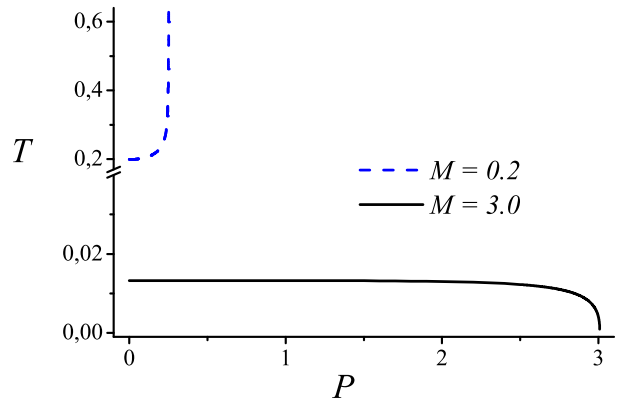}} }\hfil}%
\bigskip%
\caption{%
The $P - T$ relation.
} \label{PT}%
\end{figure}

The $M - T^{-1}$ diagram for the studied solutions is given in figure (\ref{MinvT}). Again we present the two cases $P=6.0>P_{\rm{crit}}$ and $P=0.3<P_{\rm{crit}}$. In both of them, when $M$ is large $T^{-1}\approx M$, i.e., the solutions approach the Schwarzschild black hole. The stability of the Schwarzschild black hole within the theory of Brans-Dicke has already been proved \cite{Kim} so we expect that for large $M$ our solutions are stable.
Since in both of the presented cases no turning or bifurcation point appears on the diagram we can expect that the entire equilibrium brunches are stable. With the decrease of $M$ the inverse temperature $T^{-1}$ of the solutions with $P>P_{\rm{crit}}$ increases unboundedly since an extremal solution is reached, while for $P<P_{\rm{crit}}$,  $T^{-1}$ goes to zero since an object with zero radius of the horizon, i.e. a naked singularity, is reached.
In figure (\ref{PT}) the $P - T$ relation is shown. For $M>M_{\rm{crit}}$ an extremal black hole is reached with the increase of $P$ so the temperature of the horizon goes to zero. For $M<M_{\rm{crit}}$ a naked singularity is reached with the increase of $P$ and the temperature of the horizon rises unboundedly.

\section{Stability of the solutions against spherically symmetric perturbations}
\subsection{Stability of the magnetically charged black holes}
In order to examine the stability of the solutions against radial perturbations we apply the scheme presented in details in \cite{SYT3}. First, we establish the stability against radial perturbations of the magnetically charged solutions. Then, study the electrically charged case through the electric-magnetic duality rotations.
We use the following ansatz for the metric of a spherically symmetric, time dependent space-time
\begin{equation}\label{metr}
ds^2 = - e^{\gamma}dt^2 + e^{\chi}dr^2 + e^{\beta}(d\theta^2 +
\sin^2\theta d\phi^2)
\end{equation}
where $\gamma$, $\chi$ and $\beta$ are functions of $r$ and $t$.
Then, the perturbed the fields are presented in the following way
\begin{eqnarray}
\gamma(r,t) =  \Bigl[-2\delta(r)+ \ln f(r) \Bigr]+ \triangle\gamma(r,t) ,\,\,\,\,\,\,\,\, \chi(r,t) = -\ln f(r) + \triangle\chi(r,t) , \nonumber \\
\beta(r,t) = 2\ln r + \triangle\beta(r,t) ,\,\,\,\,\,\,\, \varphi(r,t)
= \varphi(r) + \triangle\varphi(r,t),
\end{eqnarray}
where $\delta(r)$, $f(r)$ and $\varphi(r)$ give the static background solution, and $\triangle\gamma(r,t)$, $\triangle\chi(r,t)$, $\triangle\beta(r,t)$ and $\triangle\varphi(r,t)$ are small time-dependent perturbations. Let us impose the convenient gauge  $\triangle\beta(r,t)=0$. Then $e^{\beta}$ simply reduces to $r^2$.

In the case of spherically symmetric perturbations that we consider the equations decouple and the system reduces to a single equation for the perturbations of the scalar field
\begin{equation}\label{eq_phi_2}
\nabla_{\mu}^{(0)}{\nabla^{(0)}}^{\mu}\triangle\varphi -
U(r)\triangle\varphi = 0,
\end{equation}
where
\begin{multline}\label{potential}
U(r)=  -2 \Bigl\{1-2r^2\Bigl[V-{\cal A}^{4}(\varphi)L(X)\Bigr]\Bigr\}\Bigl[\partial_{r}\varphi(r)\Bigr]^2 +4r\Bigl[\partial_{r}\varphi(r)\Bigr]\frac{d V}{d\varphi}+\frac{d^2 V}{d\varphi^2}\\
+{\cal
A}^{4}(\varphi)\Bigl[16r\partial_{r}\varphi(r)\alpha(\varphi)+4\partial_{\varphi}\alpha(\varphi)\Bigr]
\Bigl[X\partial_{X}L(X) - L(X) \Bigr]  \\
-16\alpha^2(\varphi){\cal
A}^{4}(\varphi)\Bigl[X^{2}\partial^{2}_{X}L(X)-X\partial_{X}L(X)+L(X)\Bigr],
\end{multline}
and $\nabla_{\mu}^{(0)}$ is the co-derivative operator with respect to the static background.
Since the background solution is static equation (\ref{eq_phi_2}) admits a separation of the variables. Using  the following substitution
\begin{equation}\label{separ}
\triangle\varphi(r,t)=\psi(r)e^{i\omega t},
\end{equation}
from (\ref{eq_phi_2}) we obtain an equation for the spacial part of the perturbations
\begin{equation}\label{eq_psi}
f(r) e^{-\delta(r)} {d \over dr} \left[ f(r) e^{-\delta(r)} {d\psi
\over dr} \right]+{2 \over r}f^2(r) e^{-2\delta(r)}{d\psi \over
dr}+\omega^2\psi= f(r) e^{-2\delta(r)}U(r)\psi,
\end{equation}
where $\omega^2$ acts as a spectral parameter.
Equation(\ref{eq_psi}) can be cast in the form of Schr\"{o}dinger equation. In order to do this we will use the tortoise radial coordinate $r_{*}$ and will make a proper substitution for $\psi$
\begin{equation}\label{subs}
dr_{*}={dr
\over f(r) e^{-\delta(r)}},\,\,\,\,\,\,\,\,\,\,\,\,\,\,\,\,\,\psi(r)={ u(r)\over r}.
\end{equation}
The following Schr\"{o}dinger-like equation
\begin{equation}\label{Schro}
{d^{\,2} u(r_{*})\over dr_{*}^2}+\omega^2u(r_{*})=
U_{\rm {eff}}(r_{*})u(r_{*})
\end{equation}
is reached.
The effective potential is
\begin{equation}\label{GPE}
U_{\rm {eff}}(r_{*})=f(r_{*}) e^{-2\delta(r_{*})}\Bigl\{U(r_{*})-2V(\varphi)+2{\cal
A}^{4}(\varphi)L(X)+{1\over r^2(r_{*})}[1-f(r_{*})]\Bigr\}.
\end{equation}
For $r$ varying in the interval $ \left[r_{H},\infty \right)$, where $r_{H}$ is the radius of the event horizon, the tortoise radial coordinate $r_{*}\in (-\infty, \infty)$.
From this point we can use the techniques from the standard quantum mechanics to study the properties of the small perturbations.

\begin{figure}[htbp]%
\vbox{ \hfil \scalebox{1.2}{ {\includegraphics{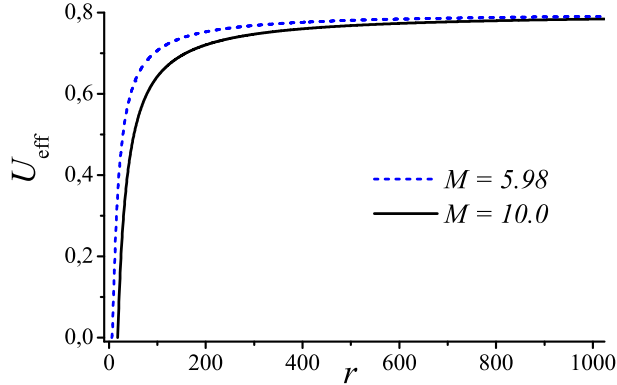}} }\hfil}%
\bigskip%
\caption{%
The effective potential $U_{\rm {eff}}$ for $\alpha=0.01$, $m_{*}^2=0.8$, $P=6.0$ and two values of the black-hole mass $M=10.0$, for which the black hole has a regular event horizon, and for $M=5.98$ corresponding to an extremal black hole.
} \label{eff_potential}%
\end{figure}

In the numerically studied region of the parameter space the effective potential $U_{\rm {eff}}$ is non-negative which means that the studied black holes are stable against spherically symmetric perturbations. Two illustrative cases are given in Figure (\ref{eff_potential}) for $\alpha=0.01$, $m_{*}^2=0.8$, $P=6.0$ and two values of the black-hole mass: $M=10.0$, for which the black hole has a regular event horizon, and $M=5.98$ corresponding to an extremal black hole.
Here we should note that in the figure the effective potential is presented in terms of the radial coordinate $r$. In terms of the tortoise coordinate $r_{*}$, $U_{\rm {eff}}$  would be more stretched but still non-negative.

\subsection{Stability of the electrically charged black holes}
The perturbations of the fields in the electrically charged case can be obtained from the magnetically charged case through the electric-magnetic duality rotations. These transformations preserve the functions of the metric and the scalar field. The perturbations of the electromagnetic field, however, will be non-vanishing in the electrically charged case. For them the duality rotations give (see \cite{SYT3})
\begin{align}
\triangle\bar{F}_{\mu\nu} = &{1\over2}(\star
G)_{\mu\nu}(\triangle\chi+\triangle\gamma)\notag\\
&+4\alpha(\varphi)X\partial^{2}_{X}L(X)(\star F)_{\mu\nu}\,\triangle\varphi-\partial_{X}L(X)(\star \triangle F)_{\mu\nu}.\label{relat}
\end{align}
Thus, the perturbations of the electromagnetic field can be expressed in terms of the functions of the background, static solution and the  perturbations of the magnetically charged solution. So the perturbations of the electrically charged solution will remain bounded with time as long as the perturbations of the magnetically charged solution are.

\section{Conclusion}

In the present work numerical solutions describing charged black holes coupled to non-linear electrodynamics in
the scalar-tensor theories with massless scalar field were found. Since an electric-magnetic duality is present
in the used electrodynamics,
only purely magnetically case was studied here. For the Lagrangian of the non-linear electrodynamics the
truncated Born-Infeld
Lagrangian was chosen and scalar-tensor theories with massive scalar field and positive coupling
parameter were considered. As a result of the
numerical and analytical investigations, some general properties of the solutions were found. The theory we considered
admits the existence of extremal black-hole solutions unlike the case with massless scalar field.

\section*{Acknowledgments} This work was partially supported by
the Bulgarian National Science Fund under Grants VUF-201/06, DO-02-257, Sofia University Research Fund No 074/2009 and Technical University – Sofia Grant 091ni143-11/2009.

\end{document}